\documentclass[reprint,amsmath,amssymb]{revtex4-1}
\usepackage[utf8x]{inputenc}
\usepackage{graphicx}% Include figure files
\usepackage{dcolumn}% Align table columns on decimal point
\usepackage{bm}% bold math
\usepackage{color} 
\usepackage{amsmath}% bold math
\usepackage{tensor}% bold mathpp

\usepackage{soul} % strikethrough
\usepackage[normalem]{ulem} % strikethrough

\begin{document}

\preprint{APS/123-QED}

\title{Fluctuations, uncertainty relations, and the geometry of quantum
  state manifolds}

\author{Bal\'azs Het\'enyi and P\'eter L\'evay}

\affiliation{MTA-BME Quantum Dynamics and Correlations Research Group,
  Department of Physics, Budapest University of Technology and
  Economics, H-1111 Budapest, Hungary}

\date{\today}% It is always \today, today,

\begin{abstract}
  The complete quantum metric of a parametrized quantum system has a
  real part (usually known as the Provost-Vallee metric) and a symplectic
  imaginary part (known as the Berry curvature).  In this paper, we
  first investigate the relation between the Riemann curvature tensor
  of the space described by the metric, and the Berry curvature, by
  explicit parallel transport of a vector in Hilbert space.  Subsequently, we write a generating function from which the complex
  metric, as well as higher order geometric tensors (affine
  connection, Riemann curvature tensor) can be obtained in terms of gauge
  invariant cumulants.  The generating function explicitly relates the
  quantities which characterize the geometry of the parameter space to
  quantum fluctuations.    We also
  show that for a mixed quantum-classical system both real and
  imaginary parts of the quantum metric contribute to the dynamics, if
  the mass tensor is Hermitian.     A many operator generalization of the
  uncertainty principle results from taking the determinant of the
  complex quantum metric.  We also calculate the quantum metric for a
  number of Lie group coherent states, including several
  representations of the $SU(1,1)$ group.  In our examples non-trivial
  complex geometry results for generalized coherent states.  A pair of
  oscillator states corresponding to the $SU(1,1)$ group gives a
  double series for its spectrum.  The two minimal uncertainty
  coherent states show trivial geometry, but, again, for generalized
  coherent states non-trivial geometry results.
\end{abstract}


\maketitle

\section{Introduction}

A parametrized quantum system exhibits interesting physics.  In such
systems a metric structure is induced on the parameter space which is
determined by the type of quantum system in question~\cite{Provost80}.
If the metric is Riemannian, it is then possible to derive an affine
connection or a Riemann curvature, quantities which further
characterize the geometry of the parameter space.  Recently, Smith et
al.~\cite{Smith22} have considered the geodesic equation associated
with a quantum metric, and arrived at equations of motion similar to
those of general relativity.  This led to the intriguing suggestion to
use parametrized quantum systems in the laboratory to mimic and study
spacetime physics.

The study of quantum metrics already has a long history.  The complex
quantum metric of a parametrized quantum system was first
derived~\cite{Provost80} by Provost and Vallee, and they showed that
its real part corresponds to a gauge invariant Riemannian metric
(known as the Provost-Vallee metric (PVM)), while its imaginary part
is of symplectic structure.  After the discovery~\cite{Berry84} of the
quantum geometric phase characterizing adiabatic cycles (also known as
the Berry phase), it was quickly realized~\cite{Shapere89} that the
imaginary part of the complex quantum metric is the Berry curvature
(BC), whose area integral within an adiabatic cycle gives the phase
itself.   \textcolor{black}{In the study of quantum phase transitions, a useful quantity is the fidelity~\cite{Sachdev11,You07,Zanardi07,Gu10,Dutta15,Carollo20,Austrich-Olivares22}, an overlap (Eq. (\ref{eqn:Sss})) between quantum states, and the fidelity susceptibility, which is a second derivative of the fidelity.   This construction is entirely equivalent to the Provost-Vallee one, which also starts from an overlap of wave functions, and in fact, the fidelity susceptibility is the PVM (Eq. (\ref{eqn:C_2_def_d}))}. The quantum metric is also relevant In the modern theory of
polarization~\cite{King-Smith93,Resta94,Resta00,Vanderbilt18}, where the
metric tensor expresses~\cite{Souza00,Ortiz00} the variance of the
polarization in insulators.  It can also be connected to
linear~\cite{Kashihara23} and non-linear~\cite{Liu23} response
functions, and can be used as a gauge to distinguish metals from
insulators~\cite{Resta05,Marrazzo19}.  More recently, the quantum
metric has reappeared in the context of topological physics,
specifically, in the description of the fractional quantum Hall
effect~\cite{Haldane11,Gromov17} and of topological
insulators~\cite{Matsuura10,Varjas22}.  Anandan and
Aharonov~\cite{Anandan90} have characterized the geometric phase by
the Fubini-Study metric, which was shown~\cite{Pati92,Levay92,Bohm03} to be
\textcolor{black}{related} to the PVM.  \textcolor{black}{The PVM is a metric on the parameter space, while the Fubini-Study metric is the metric of the projective Hilbert space of the entire system.  It follows that the PVM is the pullback of the Fubini-Study metric with respect to the map defined by the parametrized family of spectral projectors.  }  The Fubini-Study metric was also
shown~\cite{Levay04} to define a measure of entanglement.
Very recently Avdoshkin and Popov~\cite{Avdoshkin23}
  have derived quantum geometric tensors (including the Christoffel
  symbol) from a three-point Bargmann invariant~\cite{Bargmann64}.
  The Bargmann invariant can be interpreted as a cumulant generating
  function~\cite{Hetenyi22}, so the formalism of
  Ref. ~\cite{Avdoshkin23} is related to ours.

In this work we investigate the full complex quantum metric.  As a
start we investigate the relation between the two curvatures which
both appear as the characteristic of a parametrized quantum system.
The BC appears as the {\it second derivative} of the scalar product of
two quantum states with respect to the parameters, while the
four-index Riemann curvature tensor appears as the {\it fourth
  derivative}.  We clarify the relation between these two quantities
by comparing the parallel transport of a vector in Hilbert space
(which is characterized by the BC) against the parallel transport of
an ordinary vector (characterized by the Riemann curvature tensor).
Subsequently we write down the generating function from which the PVM,
as well as higher order geometric quantities can be obtained.  While
it is known that the PVM is related to the second cumulant (variance)
of the fluctuations, we point out that higher order geometric
quantities correspond to higher order cumulants, for example, the
affine connection is related to the third cumulant (skew), while the
Riemann curvature tensor corresponds to the fourth cumulant
(kurtosis).  \textcolor{black}{In the fidelity language, this means that higher order geometric tensors are related to "nonlinear fidelity susceptibilities".}

Mixed quantum-classical systems obeying the Born-Oppenheimer
approximation were considered~\cite{Mead79,Mead80} by Mead and
Truhlar.  They predicted the ``molecular Aharonov-Bohm effect'' and
introduced the notion of ``molecular magnetic field'' a gauge field
that modifies the force on the nuclei, and whose origin is the BC.
In addition, according to the analysis in Ref. \cite{Shapere89} the PVM
gives rise to a ``molecular electric'' counterpart of the molecular
magnetic field.  We revisit this issue, considering the possibility of
a complex (Hermitian) mass tensor, and find that the BC also contributes
to the molecular electric field (an extra potential term arises, since
the anti-symmetric part of the complex mass tensor couples to the BC,
which is also antisymmetric).  Complex mass was first introduced in 1967
by Feinberg~\cite{Feinberg67}.  Tachyons are quantum fields with imaginary
mass which violate physical principles and such particles are purely
hypothetical.  However, recently, Hermitian mass matrices have been
invoked to explain neutrino oscillations and generalizations of the
Dirac equation in this direction have also been put
forth~\cite{Jones-Smith14}.  By diagonalizing the inverse mass tensor,
we show that the resulting Born-Oppenheimer type system is like a "usual"
system whose only possible unusual characteristic, in addition to the
molecular fields, is that the inverse masses are directionally dependent.

Requiring that the determinant of the complex quantum metric is
greater than or equal to zero leads to a many-operator generalization
of the Schrödinger uncertainty principle~\cite{Schrodinger30}, which
is a stronger statement than the Heisenberg uncertainty principle
(whose many operator generalization was done by
Robertson~\cite{Robertson34}).  We also calculate the metric for three
types of Lie group coherent
states~\cite{Perelomov72,Gilmore72,Zhang90}: harmonic oscillator,
atomic ($SU(2)$), and hyperbolic ($SU(1,1)$).
Non-trivial complex geometry results from
  generalized~\cite{Boiteux73,Roy82} coherent states, which are
  generated from non-extremal starting states, hence they are no longer minimum
  uncertainty states.  This generalization of coherent states was
  first proposed by Boiteux and Levelut~\cite{Boiteux73}, who
  considered it a purely formal development, but later, Roy and
  Singh~\cite{Roy82} showed that the time-dependent generalization of
  these generalized coherent states lead to dynamics in which the
  probability distribution remains undistorted, and the averages
  perform classical motion.  For certain special points in the
parameter space even the quantum metric of generalized coherent states
can become trivial, points where the complex quantum metric tensor
exhibits zero determinant.  We analyze the $SU(1,1)$ case extensively,
where different representations of the group
correspond~\cite{Alhassid83,Alhassid86,Mahran86,Gerry89,Kim89,Levay93}
to different quantum number series.  A case of particular interest we
treat is a pair of coupled
oscillators~\cite{Mahran86,Gerry89,Kim89,Levay93}, which corresponds
to the direct sum of two projective representations of the $SU(1,1)$
group, corresponding to a spectrum of two series (squeezed vacuum
state, squeezed one-photon state).  In this case we find that the
complex quantum metric is zero for the extremal states of both series,
in other words, there are two cases where the geometry is trivial.

Our paper is organized as follows.  In section \ref{sec:bckgrnd} we
give a brief overview of the quantities used to characterize the
geometry of curved spaces.  In section \ref{sec:Berry} we derive the
Berry connection and curvature by explicit parallel transport, and
compare it to the affine connection and Riemann curvature.  In section
\ref{sec:cmlnts} we write down the cumulant generating function which
generates gauge invariant cumulants for a general parametrized quantum
system.  In section \ref{sec:BO_approx} we investigate a mixed
quantum-classical system with a complex mass tensor.  Then, in section \ref{sec:3index} we write the Christoffel symbol in terms of the third gauge invariant cumulant in our formalism.  In \ref{sec:fluc} we relate geometric tensors and quantum fluctuations to the uncertainty principle.  Before concluding in section \ref{sec:cnclsn}, in section \ref{sec:chrnt} we derive full quantum geometric tensors for coherent states, and identify when the geometries described by the tensors are trivial or not.

\section{Background: Geometric tensors and parallel transport}
\label{sec:bckgrnd}

In this section we provide some mathematical background~\cite{Arfken05} used in this work.  Given an $N$-dimensional metric space with coordinates $s_1,...,s_N$,
the line element squared can be written as
\begin{equation}
  dl^2 = g_{jk} d s^j d s^k,
\end{equation}
where summation over doubly occurring indices is implied, and $g_{ij}$
denotes the metric tensor.  The metric tensor characterizes the
geometry of the space and can depend on the coordinates $\{s_i\}$.

The equation for geodesic curves can be obtained by requiring that the
integral of the variance of $dl$ along a curve is zero,
\begin{equation}
  \delta \int d l = 0.
\end{equation}
This yields the second order differential equation
\begin{equation}
  g_{ki} \frac{d^2 s^i}{ d l^2} = - \frac{d s^i}{dl} \frac{d s^j}{dl} [ij,k],
\end{equation}
where $[ij,k]$ denotes the Christoffel symbol of the first kind,
\begin{equation}
  \label{eqn:Christoffel_ijk}
  [ij,k] = \frac{1}{2} \left(\partial_j g_{ik} + \partial_i g_{kj} -  \partial_k g_{ij} \right),
\end{equation}
where 
\begin{equation}
  \partial_i g_{kj} = \frac{\partial g_{kj}}{\partial s^i}.
\end{equation}
The Christoffel symbol of the second kind is defined as
\begin{equation}
  \label{eqn:Christoffel}
  \Gamma\indices{^l_i_j} = \frac{1}{2} g^{kl} (\partial_j g_{ik} + \partial_i g_{kj} -  \partial_k g_{ij}),
\end{equation}
where the inverse (or dual) of the metric tensor, $g^{ij}$ is given
by,
\begin{equation}
  g^{ij} g_{jk} = \delta\indices{^i_k}.
\end{equation}
The equation for parallel transporting a vector is usually given in
terms of the Christoffel symbol of the second kind.  If we parallel
transport a vector along a curve $s^k(\lambda)$ (parametrized by
$\lambda$) in a curved space with some metric tensor
$g_{ij}(\lambda)$, from which the Christoffel symbol is obtained
according to Eq. (\ref{eqn:Christoffel}), then the change in the
vector in the process is given by
\begin{equation}
  \label{eqn:parallel}
  \delta V^i(\lambda) = -\Gamma\indices{^i_j_k}(\lambda) V^j(\lambda) \delta s^k(\lambda).
\end{equation}

\section{Berry connection/Berry curvature and the Christoffel symbol/Riemann curvature tensor}
\label{sec:Berry}

In the case of parametrized quantum systems there are two
``connections'' and two ``curvatures''.  The Berry connection has one
coordinate index, while the affine connection associated with the
parameter space has three.  The Berry curvature is a two-index
quantity, whereas the Riemann curvature tensor has four indices.  The
Berry curvature is the imaginary part of the complex quantum metric,
which is also a two-index quantity (defined below in
Eqs. (\ref{eqn:C_2_def_a}-\ref{eqn:C_2_def_d})).   

The Berry connection and curvature correspond~\cite{Simon83} to the
gauge fields of a fiber bundle.   \textcolor{black}{A fiber bundle is constructed by first considering the projected Hilbert space of the parametrized quantum system, where the projection occurs onto the space spanned by a chosen wave function or chosen set of wave functions.  The projection removes the phase indeterminacy (for example, if a single state is used, then the projection, $|\Psi(s)\rangle \langle \Psi(s) |$, has no arbitrary phase due to the simultaneous presence of a bra and a ket).   A fiber bundle restores the phase by assigning a fiber to each point of the base space (which in this case is the parameter space $s$).   For example, in the case of a single state projection, a typical fiber can be the space on which the phase is easily represented, the $S^1$ unit circle.}
In this sense, there is a difference between the Berry connection/curvature (which corresponds to the gauge fields which live on the fibers) and the Christoffel and Riemann curvature tensors (which live on the parameter space $s$), however, we demonstrate 
below a useful parallel between these two sets of quantities.

We will argue below that the above discrepancy between the number of indices between the two sets of quantities is due
  to the projection of the Hilbert space necessary to obtain
  physically interesting cases (it is conventional to suppress the indices of the projected states).  It is in order to cite a few examples.  The Berry phase is defined via
  an integral of one quantum state over an adiabatic cycle, and the
  state index is usually not explicitly indicated when writing the Berry
  connection.  The state under scrutiny is separated by an energy gap
  from the other states.  A sum over Berry phases over all the states
  of a complete Hilbert space is zero~\cite{Pacher89,Xiao10}.  In the
  case of the generalization of the Berry phase to a degenerate subset
  of states by Wilczek and Zee~\cite{Wilczek84}, it is the subset of states that is
  carried around an adiabatic cycle, again, separated from the rest of
  the states by a gap.  In the modern theory of polarization of
  insulating systems~\cite{King-Smith93,Resta94,Resta00,Vanderbilt18},
  a particular kind of Berry phase, a Zak phase~\cite{Zak89}, is
  evaluated, which is an integral over the Brillouin zone over an
  occupied band (or bands) of states.  The projection here is justified by
  the fact that only some bands are occupied, unoccupied bands do not
  contribute.  In the case of time-reversal invariant topological
  insulators~\cite{Kane05a,Kane05b} a modified version of the Zak
  phase is calculated for occupied degenerate Kramers bands.  While
  the gap condition can be relaxed~\cite{Aharonov87}, a non-trivial
  value for a Berry phase or a Wilson loop still requires a projection in
  Hilbert space.

We now turn to the main purpose of this section, which is to derive
the Berry connection and curvature by explicit parallel transport of a
vector in Hilbert space.  Our aim is to place emphasis on the
connections (no pun intended) and the differences between the Berry
connection and curvature on the one hand and the Christoffel symbol
and the Riemann curvature tensor on the other.

One way~\cite{Arfken05} to arrive at the Christoffel symbol and the Riemann tensor is
to consider the change in a general vector ${\bf V}(s)$ defined on a curved parameter space.   \textcolor{black}{${\bf V}(s)$ is a mapping from each point of the parameter space $s_1,...,s_N$ to $V_1,...,V_M$.   One can expand the vector in a local basis,}
\begin{equation}
  {\bf V}(s) = V^j(s) {\bf e}_j(s),
\end{equation}
where $V^j(s)$ denote the components of the vector ${\bf V}(s)$ in the local basis and ${\bf
  e}_j(s)$ denote the members of the basis themselves.  In general, a change in
${\bf V}(s)$ results from a change in the coefficients themselves as well as a possible change in the
basis,
\begin{equation}
  \frac{\partial{\bf V}(s)}{\partial s^k} = \frac{\partial {V^j(s)}}{\partial s^k} {\bf e}_j(s) + V^j(s) \frac{\partial  {\bf e}_j(s)}{\partial s^k}.
\end{equation}
We can expand the derivative in the second term as,
\begin{equation}
  \frac{\partial{\bf e}_j(s)}{\partial s^k} = \Gamma\indices{^l_j_k}(s) {\bf e}_l(s),
\end{equation}
where the Christoffel symbol, $\Gamma\indices{^l_j_k}(s)$ is the
expansion coefficient.  Let us resolve the change in ${\bf V}(s)$ into
components as
\begin{equation}
  \label{eqn:partial_V}
  {\bf e}^l(s) \cdot \frac{\partial{\bf V}(s)}{\partial s^k} = \frac{\partial {V^l(s)}}{\partial s^k} + V^j(s) \Gamma\indices{^l_j_k}(s).
\end{equation}
We define the covariant derivative in the $k$ direction as,
\begin{equation}
  D_k V^l(s) = \frac{\partial {V^l(s)}}{\partial s^k} + \Gamma\indices{^l_j_k}(s)V^j(s).
\end{equation}
From the covariant derivative, it is possible to define the four-index
Riemann curvature tensor, as
\begin{equation}
  \label{eqn:DD}
  (D_j D_k - D_k D_j)V^l(s) = R\indices{^l_{m}_k_j}(s)V^m(s),
\end{equation}
where
\begin{eqnarray}
  \label{eqn:R}
  R\indices{^l_{m}_k_j}(s) &=& \partial_j\Gamma\indices{^l_k_m}(s) - \partial_k\Gamma\indices{^l_j_m}(s) \\ & &+ \Gamma\indices{^l_j_n}(s)\Gamma\indices{^n_k_m}(s) - \Gamma\indices{^l_k_n}(s)\Gamma\indices{^n_j_m}(s). 
  \nonumber
\end{eqnarray}
Note, the first pair of indices in $R\indices{^l_{m}_k_j}$ ($l$ and
$m$) rotate the vector components, while the second pair ($k$ and $j$)
are sensitive to the curvature of the surface on which the transport
occurs.

In an exactly analogous manner, we can consider a quantum system which
is parametrized by a set of variables $s = s_1,...,s_N$, and we can
write a general vector in the Hilbert space in some basis,
\begin{equation}
  | \Psi(s) \rangle = \sum_\iota |\iota(s)\rangle \langle \iota(s) |
  \Psi(s) \rangle = \sum_\iota |\iota(s)\rangle \Psi^\iota(s).
\end{equation}
We use Greek indices to label basis states.  We can consider a change
in this state in the $k$-direction,
\begin{equation}
  \partial_k | \Psi(s) \rangle = \sum_\iota \left[ |\iota(s)\rangle \partial_k \Psi^\iota(s)
    + \partial_k |\iota(s)\rangle \Psi^\iota(s) \right].
\end{equation}
For the directions in the parameter space we use Latin indices.
Multiplying from the left by another state $\langle \lambda(s) |$ we
obtain the change of $\Psi(s)$ in the basis,
\begin{eqnarray}
  \label{eqn:nabla_Psi}
  \tilde{D}_k \Psi^\lambda &=& \langle \lambda(s) |\partial_k | \Psi(s) \rangle \\
   &=& \partial_k \Psi^\lambda(s) +
  \sum_\iota \langle \lambda(s) | \partial_k | \iota(s) \rangle \Psi^\iota(s).\nonumber
\end{eqnarray}

Comparing Eqs. (\ref{eqn:partial_V}) and (\ref{eqn:nabla_Psi}), we see
that the two terms on the right hand side of the equations correspond.
The analog of the Christoffel symbol in Eq. (\ref{eqn:nabla_Psi}) is
the expression
\begin{equation}
  \tilde{\Gamma}\indices{^\lambda_{\iota}_k} = \langle \lambda(s) | \partial_k | \iota(s) \rangle.
\end{equation}

The Berry connection is a special case of
$\tilde{\Gamma}\indices{^\lambda_{\iota}_k}$.  The Berry phase arises
from transporting a {\it single} energy eigenstate around an adiabatic
cycle.  In other words, the state $|\Psi(s)\rangle$ is itself some
energy eigenstate, and the basis used is also the energy eigenbasis,
resulting in
\begin{equation}
  i \tilde{\Gamma}\indices{^0_{0}_k} = i \langle \Psi_0(s) | \partial_k
  | \Psi_0(s) \rangle.
\end{equation}
The general connection, $\tilde{\Gamma}\indices{^\lambda_{\iota}_k}$, is
a three-index quantity because it connects the states indexed by
$\lambda$ and $\iota$ by a change in the coordinate $s_k$.
$\tilde{\Gamma}\indices{^\lambda_{\iota}_k}$ represents the probability
amplitude to go between quantum state $\lambda$ and $\iota$ when a
change in the coordinate $k$ is occurring.  The Berry connection
connects the same states, and this state remains fixed during the
entire adiabatic cycle, and the only remaining index is the coordinate
one, $k$.   When $\lambda \neq \iota$ , and the system is non-degenerate, the quantity $\tilde{\Gamma}\indices{^\lambda_\iota_k}$ represents non-adiabatic transitions between different quantum states under a change of the coordinate $s_k$.    As such, this quantity plays a central role in the field of non-adiabatic molecular dynamics~\cite{Tully90,Webster91,Coker93}: the non-adiabatic coupling vector is closely related~\cite{Coker95} to the Pechukas~\cite{Pechukas69a,Pechukas69b} force, the force which acts on the nuclei of a molecular system whose electrons undergo a transition between electronic quantum states, in other words, when the Born-Oppenheimer approximation does not hold.

We can also derive the analog of the Riemann curvature tensor in this
case, via
\begin{equation}
  \label{eqn:tld_DD}
  (\tilde{D}_k \tilde{D}_l - \tilde{D}_l \tilde{D}_k)\Psi^\lambda =
  \tilde{R}\indices{^\lambda_{\mu}_l_k}\Psi^\mu,
\end{equation}
resulting in
\begin{equation}
  \label{eqn:tld_R}
  \tilde{R}\indices{^\lambda_{\mu}_l_k} =
  \partial_k \tilde{\Gamma}\indices{^\lambda_{\mu}_l} -
  \partial_l \tilde{\Gamma}\indices{^\lambda_{\mu}_k} +
  \tilde{\Gamma}\indices{^\lambda_{\nu}_k} \tilde{\Gamma}\indices{^\nu_{\mu}_l}  -
  \tilde{\Gamma}\indices{^\lambda_{\nu}_l} \tilde{\Gamma}\indices{^\nu_{\mu}_k}.
\end{equation}
\textcolor{black}{Eq. (\ref{eqn:tld_R}) is identically zero if the indices run over the entire Hilbert space.  To show this, we write the $s$-dependent state $|\iota(s)\rangle$ using a fixed basis, as, 
\begin{equation}
|\iota(s) = U(s) |\iota_0 \rangle,
\end{equation}
where $|\iota_0 \rangle$ denotes a member of the fixed basis, and $U(s)$ is a unitary matrix spanning the entire Hilbert space.   The quantum Christoffel symbol becomes,
\begin{equation}
  \tilde{\Gamma}\indices{^\lambda_{\iota}_k} = \langle \lambda_0 | U(s)^\dagger \partial_k U(s) | \iota_0 \rangle.
\end{equation}
Using this definition of $\tilde{\Gamma}\indices{^\lambda_{\iota}_k} $ it is a relatively simple exercise that $\tilde{R}\indices{^\lambda_{\mu}_l_k}$, as defined in Eq. (\ref{eqn:tld_R}), is zero.  But this only holds if the entire Hilbert space is considered, in other words, a non-trivial quantum Riemann curvature results if the Hilbert space is truncated, meaning that all the indices in Eq. (\ref{eqn:tld_R}) (including the internal ones which are summed) are truncated~\cite{Pacher89}.}    Projections of the types discussed above are examples of such truncations.   Comparing the sets
of Eqs.  (\ref{eqn:DD}), (\ref{eqn:R}) and (\ref{eqn:tld_DD}),
(\ref{eqn:tld_R}), we see that there is a correspondence between the
curvature tensor of the parametrized quantum system and the Riemann
curvature tensor known from differential geometry.

The four-index tensor $\tilde{R}\indices{^\lambda_{\mu}_l_k}$ is not
a curvature in the sense of the Riemann curvature tensor.  To lower
the index $\lambda$ in $\tilde{R}\indices{^\lambda_{\mu}_l_k}$ one
uses not the metric tensor of the parameter space, but that of the
Hilbert space of states ($\delta_{\mu \nu}$ for complete orthonormal
states).  As for contraction the Riemann curvature tensor can be
contracted in three ways, because the indices are equivalent, whereas
$\tilde{R}\indices{^\lambda_{\mu}_l_k}$ can only be contracted as
\begin{equation}
  \tilde{R}\indices{_l_k} = \tilde{R}\indices{^\lambda_{\lambda}_l_k}.
\end{equation}
The contraction over the entire Hilbert space leads to zero (for a
proof of this statement, see for example, the introduction of
Ref. \onlinecite{Xiao10}), however it is possible to sum over a part
of the Hilbert space.  The Berry curvature involves only one state,
its generalization~\cite{Wilczek84}, the Wilson loop involves a subset
of the complete Hilbert space (usually a degenerate subspace).  In
these cases $\tilde{R}\indices{_l_k}$ is not trivial.

If the metric tensor of the parameter space is invertible, one can
raise one of the indices of $\tilde{R}\indices{_l_k}$ as
\begin{equation}
  \tilde{R}\indices{^j_k} = g\indices{^j^l}\tilde{R}\indices{_l_k},
\end{equation}
and contract it to get the scalar curvature,
\begin{equation}
  \tilde{R} = \tilde{R}\indices{^j_j}.
\end{equation}
The scalar curvature obtained from the Berry curvature is zero on
account of the anti-symmetry in the two coordinate indices.

\section{Cumulant generating function for quantum states}
\label{sec:cmlnts}

Before stating the cumulant generating function relevant to quantum
state manifolds, let us give the definitions of moments and cumulants
from the theory of probability.  In this section, we use a notation that is closest to the original derivation of Provost and Vallee~\cite{Provost80}.   In the Appendix we develop a notation in which the relation between moments and cumulants in the case of quantum state manifolds can be made manifest.

Moments and cumulants are quantities which characterize probability
distributions.  Given a multi-variate normalized probability
distribution:
\begin{eqnarray}
  P(x_1,...,x_N) \geq 0, \forall x_i; \hspace{3cm}\\
  \int_{-\infty}^\infty...\int_{-\infty}^\infty d x_1 ... d x_N P(x_1,...,x_N) = 1, \nonumber
\end{eqnarray}
the $M$th moment and the $M$th cumulant is obtained from the
generating (or characteristic) function, defined as,
\begin{equation}
  f(k_1,...,k_N) = \int d x_1 ... d x_N e^{i \sum_{j=1}^N k_j x_j} P(x_1,...,x_N),
\end{equation}
via the derivatives and logarithmic derivatives, respectively, as
\begin{eqnarray}
  \label{eqn:MC}
  \mathcal{M}_M &=&  \left. \prod_{j=1}^N \frac{1}{i^{m_j}}\frac{\partial^{m_j}}{\partial k_j^{m_j}} f(k_1,...,k_N) \right|_{k_1=...=k_N=0}, \\ \nonumber
  \mathcal{C}_M &=& \left. \prod_{j=1}^N \frac{1}{i^{m_j}}\frac{\partial^{m_j}}{\partial k_j^{m_j}} \ln f(k_1,...,k_N) \right|_{k_1=...=k_N=0},
\end{eqnarray}
where $\sum_{j=1}^N m_j = M$.  Cumulants can be written in terms of
moments (and vice versa).

We now turn to quantum state manifolds.  Consider a family $\{ \Psi(s)
\}$ of quantum state vectors parametrized smoothly by the
$n$-dimensional variable $s = s_1,...,s_n$.  We write the overlap
between two wave functions with different $s$ as,
\begin{equation}
  \label{eqn:Sss}
S(s',s) =  \langle \Psi(s')| \Psi(s) \rangle. 
\end{equation}
We define the generating function for gauge invariant cumulants as
\begin{eqnarray}
  \label{eqn:Cmlnts}
  \mathcal{C}_{m_1',...,m_N';m_1,...,m_N} =  \hspace{2.5cm} \\ 
   (-i)^{M'}(i)^M (\partial_1')^{m_1'} ...(\partial_N')^{m_N'} 
  \partial^{m_1}_1...\partial^{m_N}_N
  \left. \ln S(s',s) \right|_{s' = s}. \nonumber
\end{eqnarray}
In Eq. (\ref{eqn:Cmlnts}) $M' = m_1' + ... + m_N'$ and $M = m_1 +
... + m_N$, and $\partial_j$($\partial_j'$) denotes a partial
derivative with respect to $s_j$($s_j'$).  Since this cumulant depends
on the parameters of the vector in Hilbert space $| \Psi(s) \rangle$
and its dual $\langle \Psi(s')|$, there are two numbers specifying the
order of the cumulant, $M'$, referring to the dual vector $\langle
\Psi(s')|$ and $M$ referring to the vector $|\Psi(s) \rangle$.  In
Section \ref{sec:fluc} we show that in the case where the state
$|\Psi(s)\rangle$ is generated by simple translation operators,
$\mathcal{C}_{m_1',...,m_N';m_1,...,m_N}$ correspond exactly to the
statistical cumulants given in Eq. (\ref{eqn:MC}).

Gauge invariance can be shown via substituting a wave function
\begin{equation}
  |\tilde{\Psi}(s)\rangle = \exp(-i \alpha(s))|\Psi(s)\rangle,
\end{equation}
a state with a modified phase, $\alpha(s)$ which depends on the
parameter set $s$.  The overlap becomes,
\begin{equation}
  \label{eqn:S_ss}
  \tilde{S}(s',s) = \ln \langle \tilde{\Psi}(s') | \tilde{\Psi}(s) \rangle = \ln \langle \Psi(s') | \Psi(s) \rangle + i (\alpha(s') - \alpha(s)).
\end{equation}
While cumulants with either $M$ or $M'$ zero are not gauge invariant,
all other cases (including cumulants which correspond to the geometric
tensors of interest) are independent of $\alpha(s)$.

The gauge invariant quantum metric tensor of Provost and
Vallee~\cite{Provost80} is easily generated.  We first relabel the
cumulants of order $M=1$ and $M'=1$ as
\begin{equation}
  \label{eqn:C_2_def_a}
  C_2(j;k) = C_{0,...,0,m_j'=1,0,....,0;0,...,0,m_k=1,0,....,0}.
\end{equation}
Evaluating $C_2(j;k)$ using Eq. (\ref{eqn:Cmlnts}), we obtain
\begin{equation}
  \label{eqn:C_2_def_b}
  C_2(j;k) = \gamma_{jk} - \beta_j \beta_k,
\end{equation}
where
\begin{eqnarray}
  \label{eqn:C_2_def_c}
  \gamma_{jk} &=& \langle \partial_j \Psi(s) | \partial_k\Psi(s)\rangle, \\
  \beta_j &=& -i \langle \Psi(s) | \partial_j\Psi(s)\rangle. \nonumber
\end{eqnarray}
$\beta_j$ is the Berry connection in the state $\Psi(s)$.  The metric
tensor $g_{jk}$ and the Berry curvature $\sigma_{jk}$ are given by
\begin{equation}
  \label{eqn:C_2_def_d}
  g_{jk}(s) = \mbox{Re} C_2(j;k);\hspace{1cm} \sigma_{jk}(s) = \mbox{Im} C_2(j;k).
\end{equation}
$g_{jk}$($\sigma_{jk}$) is symmetric(anti-symmetric) in its indices.
The second cumulant in ordinary statistics gives only a real
correlation matrix.  In the quantum case the particular structure of
the scalar product of the probability amplitudes of quantum states
($S(s',s)$) is what gives rise to the additional imaginary
anti-symmetric component.

\section{A Born-Oppenheimer system with Hermitian inverse mass tensor}

\label{sec:BO_approx}

One area in which the study of the PVM and the BC
  proved crucially important was for mixed quantum-classical systems,
  in which the quantum subsystem remains in one quantum state
  throughout the dynamics (Born-Oppenheimer approximation).  Mead and
  Truhlar showed that the BC gives rise to the ``molecular magnetic
  field'', which is measurable in several types of experiments, for
  example, pseudorotation of triatomic
  molecules~\cite{Delacetraz86,Wolf89}.

In this section we show, on the one hand, that the metric tensor as
well as its derivative (a three-index quantity related to the
Christoffel symbol) appears in mixed quantum classical systems.
In addition, we also show that a complex inverse mass
  tensor leads to a ``molecular electric field'' which depends not
  only on the PVM, but also the BC.

We consider a molecular system consisting of electrons and nuclei.
Since nuclei are at least three orders of magnitude heavier than
electrons, the dynamics of molecules are often studied by invoking the
Born-Oppenheimer approximation, which assumes that electrons remain in
one electronic state (the ground state).  We first write the
Hamiltonian (fully quantum) as
\begin{equation}
  \label{eqn:H_full}
  \hat{H}_{tot} = \frac{1}{2} \sum_{jk} Q_{jk}^* \hat{P}_j \hat{P}_k +
  \hat{h}(\hat{\xi};\hat{s}),
\end{equation}
where $\hat{s}_j$ and $\hat{P}_j = i \partial_{s_j}$ represent the
positions and momenta of the nuclei, \textcolor{black}{$\hat{\xi}$ denote the
electronic coordinates collectively ($s$ denotes the nuclear coordinates collectively)}, and
$h(\hat{\xi};\hat{s})$ is the Hamiltonian of the electrons.
$Q_{ij}^*$ denotes the complex conjugate of the inverse mass tensor.
We assume that the inverse mass tensor has real and imaginary parts,
\begin{equation}
  Q_{jk} = Q'_{jk} + i Q''_{jk}.
\end{equation}
From the hermiticity of the kinetic energy, it follows that $Q'_{jk} =
Q'_{kj}$, and $Q''_{jk} = - Q''_{kj}$.

In the Born-Oppenheimer approximation one writes the total wave
function as:
\begin{equation}
  \Psi_{tot}(\xi;s) = \psi_{nuc}(s)\phi_0(\xi;s),
\end{equation}
where $\psi_{nuc}(\hat{X})$ denotes the nuclear wave function, and
$\phi_0(\hat{\xi};\hat{X})$ denotes the ground state electronic wave
function, satisfying
\begin{equation}
  \hat{h}(\xi;s) \phi_0(\xi;s) = E_0(s)\phi_0(\xi;s).
\end{equation}
The effective Hamiltonian acting on the nuclei that results is
\begin{equation}
  \hat{H}_{eff} = \int d {\xi} \phi_0^*(\xi;s)\hat{H}_{tot}
  \phi_0(\xi;s).
\end{equation}
The effective Hamiltonian can be shown to be
\begin{eqnarray}
  \nonumber
  \hat{H}_{eff} &=&  \sum_{jk} \frac{Q_{jk}^*}{2} ( \hat{P}_j -
  A_j(s)) ( \hat{P}_k - A_k(s)) \\ & & + \Phi(s) + E_0(s),
  \label{eqn:Heff}
\end{eqnarray}
where, $A_j(s)$ and $\Phi(s)$ denote a vector potential
and a scalar potential, respectively, defined as:
\begin{eqnarray}
  \label{eqn:APhi}
  A_j(s) &=& - i \int d {\bf \xi}
  \phi_0^*(\xi;s)\partial_j \phi_0(\xi;s)
  \\ \nonumber \Phi(s) &=& \frac{1}{2}\sum_{jk} Q_{jk}^*
  C_2(j;k) 
  \\ \nonumber &=& \frac{1}{2}\sum_{jk} (Q'_{jk}g_{jk}(s)+Q''_{jk}\sigma_{jk}(s)).
\end{eqnarray}
$g_{jk}(s)$ denotes the quantum metric tensor, derived in the previous
section.  In Eq. (\ref{eqn:APhi}) the real part of the mass tensor
couples to the PVM, and the imaginary part couples to the BC.  Even
though the mass tensor has an imaginary component, the terms in the
effective Hamiltonian obtained above are all real, so the above
potential corresponds to a classical system with a modified potential
(modified ``molecular electric field''). 

It is also possible to diagonalize the Hermitian mass
  tensor, and end up with a more straightforward expression:
  $\hat{H}_{eff}$, 
\begin{equation}
  \hat{H}_{eff} =  \sum_{r} \frac{\hat{\Pi}_r^2}{2 M_r} + \Phi(s) + E_0(s),
  \label{eqn:Heff_diagQ}
\end{equation}
where
\begin{equation}
  \hat{\Pi}_r = \sum_j U_{jr} ( \hat{P}_j - A_j(s)),
\end{equation}
where $U$ diagonalizes $Q$, and $M_r$ are the inverse eigenvalues.
The potential adopts the form,
\begin{equation}
  \Phi(s) = \sum_r \frac{\tilde{g}_r(s) + \tilde{\sigma}_r(s)}{M_r},
\end{equation}
where
\begin{eqnarray}
  \tilde{g}_r(s) &=& \sum_{jk} (U_{jr}U_{kr}^*)' g_{jk}(s)\\
  \tilde{\sigma}_r(s) &=& \sum_{jk} (U_{jr}U_{kr}^*)'' \sigma_{jk}(s). \nonumber
\end{eqnarray}
In the form of Eq. (\ref{eqn:Heff_diagQ}) the
  effective Hamiltonian is not very different from systems with an
  ``effective mass'', since the kinetic energy has the usual from,
  apart from exhibiting directionally dependent masses, and the BC term only effects the
  potential in which the particle is moving.  It is interesting that
  in the original Hamiltonian (Eq. (\ref{eqn:H_full})), the imaginary
  part of the complex Hermitian mass tensor appears in the kinetic
  energy, however, after invoking the Born-Oppenheimer approximation,
  it only effects the potential energy.
  
\section{Three index quantities: The quantum Christoffel symbol}
\label{sec:3index}

We now consider the three-index quantity of the form
\begin{equation}
\label{eqn:C3_jkl}  
  C_3(j;kl) = (-i\partial_{j'})(i\partial_{k})(i\partial_{l})
  \left. \ln S(s',s) \right|_{s' = s}.
\end{equation}
$C_3(j;kl)$, like $C_2(j;k)$, is complex and takes the form,
\begin{eqnarray}
  \label{eqn:C3_jkl_2}  
  C_3(j;kl) &=& i \langle \partial_j \Psi |\partial^2_{kl} \Psi
  \rangle - \langle \Psi |\partial^2_{kl} \Psi \rangle \beta_j \\ & &
  + \gamma_{jl} \beta_k + \gamma_{jk} \beta_l - 2 \beta_j \beta_k
  \beta_l. \nonumber
\end{eqnarray}
It is possible to show that
\begin{equation}
  \label{eqn:dlC2jk}
  \partial_l C_2(j;k) = \frac{1}{2i} \left(C_3(jl;k) - C_3(j;kl)\right),
\end{equation}
by directly taking the derivative of $C_2(j;k)$ (as
  defined in Eq. (\ref{eqn:C_2_def_b})) with respect to $s_l$, and
  substituting Eq. (\ref{eqn:C3_jkl_2}) in the right hand side of
  Eq. (\ref{eqn:dlC2jk}).  Using this result, we now define the
quantum extension of the Christoffel symbol as,
\begin{eqnarray}
  \label{eqn:QChrstffl}
  [jl;k]_q &=& \frac{1}{2}\left( \partial_j C_2(l;k) + \partial_l C_2(k;j)
  - \partial_k C_2(j;l) \right),\\
&=& \frac{1}{4i}\left(C_3(jl;k) - C_3(l;kj) + C_3(lk;j) - C_3(k;jl)\right. \nonumber \\
&& \left. - C_3(kj;l) + C_3(j;lk) \right). \nonumber
\end{eqnarray}
The quantum Christoffel symbol is the analog of the expression in Eq. (\ref{eqn:Christoffel_ijk}), with the full quantum metric substituted in place of $g_{jk}$.   In addition, the quantum Christoffel symbol evaluates to a sum of various
  third order cumulants.  Eq. (\ref{eqn:QChrstffl}) establishes a
  definite relation between quantum fluctuations and the Christoffel
  symbol.

The usual Christoffel symbol of the first kind, which results from
taking the derivative of the PVM, is given by
\begin{equation}
  \mbox{Re}[jl;k]_q = [jl;k],
\end{equation}
and
\begin{equation}
  \mbox{Im}[jl;k]_q = \frac{1}{2}\left( \partial_j \sigma_{lk} + \partial_l \sigma_{kj}  - \partial_k \sigma_{jl} \right).
\end{equation}
In the usual Christoffel symbol (Eq. (\ref{eqn:Christoffel})), the
first and the last terms change sign upon exchanging indices $j$ and
$k$, and in $\mbox{Im}[jl;k]_q$ exactly the opposite happens.  For
this reason, using $[jl;k]_q$, we can write the force associated with
the potential $\Phi(\hat{X})$ as
\begin{eqnarray}
  F_l &=& -\frac{1}{2}\sum_{jk} Q_{jk}^* [jl;k]_q, \\
      & & -\frac{1}{2}\sum_{jk} \left( Q'_{jk} \mbox{Re}[jl;k]_q + Q''_{jk} \mbox{Im}[jl;k]_q\right).\nonumber
\end{eqnarray}
The force is thus dependent on the extended quantum Christoffel
symbol.  The broader conclusion is that in the
  most general case the parallel transport of a vector in the
  parameter space $s$ proceeds via not only the derivative of the PVM,
  but also of the BC.  A quantum Christoffel symbol of the second
kind can only be defined if the complex quantum metric $C_2(j;k)$ is
invertible.  In that case the definition in
Eq. (\ref{eqn:Christoffel}) can be used.

Since it is always possible to take further derivatives according to
Eq. (\ref{eqn:dlC2jk}), and the resulting higher order cumulants can
then be combined to conform to expressions of higher order geometric
tensors (for example, the Riemann curvature tensor), geometric tensors
of any order can be expressed as quantum fluctuations of the same
order.

\section{Geometric tensors, quantum fluctuations, and the uncertainty principle}
\label{sec:fluc}

In this section we investigate how quantum fluctuations are related to
geometric tensors.  The most straightforward example~\cite{Provost80}
is states generated by commuting operators $\{\hat{A}_j\}$, from some
quantum state $| \Psi_0 \rangle$, as
\begin{equation}
  \label{eqn:Psi_clsscl}
|\Psi \rangle = \exp \left( i \sum_j s_j \hat{A}_j\right) | \Psi_0\rangle.
\end{equation}
Apart from the fluctuations in $| \Psi_0 \rangle$, this example can be
considered classical.  Substituting Eq. (\ref{eqn:Psi_clsscl}) into
Eq. (\ref{eqn:Cmlnts}), we obtain ordinary statistical cumulants
(Eq. \ref{eqn:MC}).  The metric tensor $g_{jk}$ is simply
the second statistical cumulant, $\sigma_{jk}$ is zero in this case.
$C_3(jk;l)$ only has a real part and corresponds to the third order
statistical cumulant (the skew).

Further insight can be gained via considering a state generated by
assuming that the set of operators $\{\hat{A}_j\}$ are non-commuting
(generators of a Lie group).  To start, we study the fluctuations near
the origin of the parameter space, by taking $s$ to zero in
Eq. (\ref{eqn:Cmlnts}).  In this case, the cumulants will have
imaginary parts.  The second order cumulant looks like,
\begin{eqnarray}
\nonumber  g_{jk} &=& \frac{1}{2}\langle[\hat{A}_j,\hat{A}_k]_+ \rangle_0 - \langle \hat{A}_j \rangle_0 \langle \hat{A}_k \rangle_0, \\ 
\sigma_{jk} &=& \frac{1}{2}\langle [\hat{A}_j,\hat{A}_k]_- \rangle_0,
\label{eqn:C2_0}
\end{eqnarray}
where $[,]_+$($[,]_-$) indicates an anti-commutator(commutator), and
$\langle\rangle_0$ indicates average over the state $|\Psi_0\rangle$.
A non-trivial Berry phase is a result of non-commuting generators.

It is instructive to consider the matrix $C_2(j;k)$ for a system with
a two-dimensional parameter space, where there are only two
generators, $\hat{A}_1$ and $\hat{A}_2$.  In this case, the quantities
$C_2(1;1)$ and $C_2(2;2)$ are simply the variances of operators
$\hat{A}_1$ and $\hat{A}_2$.  Since the quantity $C_2(j;k)$ is itself
a variance, and therefore has a determinant greater than or equal to
zero, we obtain 
\begin{equation}
  \label{eqn:uncertainty_2}
  \sigma_j^2 \sigma_k^2 \geq \left| \frac{1}{2} \langle [\hat{A}_j,\hat{A}_k]_+
  \rangle_0 - \langle \hat{A}_j \rangle_0 \langle \hat{A}_k \rangle_0 \right|^2 -
\left| \frac{1}{2}\langle [\hat{A}_j,\hat{A}_k]_- \rangle_0 \right|^2.
\end{equation}
This equation is a form of the uncertainty relation, known as the
Schr\"odinger uncertainty relation~\cite{Schrodinger30}, which is a
stronger form of this principle than the well-known one due to
Heisenberg.  Considering now the general case of an arbitrary number
of dimensions, a multi-dimensional generalization of the uncertainty
principle can be obtained via
\begin{equation}
  \label{eqn:uncertainty_N}
  \mbox{det} [C_2(j;k)] \geq 0.
\end{equation}
This form of the uncertainty principle places a constraint on the
variances $\sigma_1, ... \sigma_N$ if the parameter space is $N$
dimensional.  The many-operator version of the Heisenberg uncertainty
relation was first derived by Robertson~\cite{Robertson34}.

If the determinant of $C_2(j;k)$ is zero, it means that the complex
metric $C_2(j;k)$ can not be inverted.  This also means that a
Christoffel symbol of the second kind can not be constructed, and that
the parameter space, when considered together with the quantum
fluctuations, has a trivial geometry, the entire parameter space
represents the same point (all distances are zero).  In the next
section we give examples of both cases.

\section{Coherent state examples}

\label{sec:chrnt}

\textcolor{black}{In this section we will analyze the geometry of quantum systems using coherent states.  Reviews of this subject are found in Refs. \onlinecite{Zhang90,Gilmore72,Perelomov86}.   After deriving the uncertainty relation in a coherent state context, we will consider three well-known examples of coherent states associated with Lie groups.  They are known as Glauber, $SU(2)$ and $SU(1,1)$ coherent states.   Glauber coherent states are associated with the Weyl-Heisenberg group~\cite{Weyl50}.   Glauber coherent states are minimum uncertainty states of a quantum harmonic oscillator.  $SU(2)$ coherent states are constructed using angular momentum operators.  The generators of the $SU(1,1)$ group can be obtained by modifications of the $SU(2)$ algebra.}   \textcolor{black}{An interesting connection~\cite{Alhassid83,Alhassid86} exists between the $SU(2)$ and $SU(1,1)$ groups.   The $SU(2)$ space corresponds to a sphere, which is a compact space.  The basis states of the $SU(2)$ correspond to the bound states of the P\"oschl-Teller potential.  The modifications leading to the $SU(1,1)$ algebra lead to a non-compact group space (a hyperboloid) and the modified basis states correspond to the scattering states of the same potential.   The $SU(1,1)$ group space has several projections, which correspond to different series of quantum numbers.   In the case of $SU(1,1)$ we will consider several representations of the group space below, including the case when the group space is the universal covering space of the hyperboloid, as well as projections thereof.}

Before turning to our concrete examples, we give a brief overview of
coherent states.    \textcolor{black}{Our purpose here is to calculate and analyze the quantum geometric tensor associated with coherent states, but we believe it is in order to give the general steps of their construction in a Lie group context.}

Given a Lie group $G$ with a Lie algebra
consisting of the identity $\hat{I}$ and two sets of generators,
$\{\hat{A_i}\}$ and $\{\hat{B_i}\}$.   The original coherent states~\cite{Zhang90} are minimum uncertainty states whose starting point is an {\it extremal}
state.  \textcolor{black}{The set $\{\hat{A_i}\}$ together with the identity forms the stability subgroup of $G$, which we call $H$, consisting of
operators which leave the extremal state invariant.    The set $\{\hat{B_i}\}$, the complement of $H$ in $G$ consists of operators which do not leave the extremal state invariant, and include at least one operator, which eliminates the extremal state.  In practice, this means that the extremal state is an eigenstate of the operators included in $H$, with definite quantum numbers, and the operators of the  complement of $H$ in $G$ are ladder (raising and lowering, or creation and annihilation) operators.}  Coherent states are then constructed by applying the shift operator, $\hat{D}(s)$, 
\begin{equation}
  \label{eqn:cs_Glauber}
  |s \rangle = \hat{D}(s) |0\rangle, \hat{D}(s) = \exp\left( \sum_j s_j
  \hat{B_j}\right),
\end{equation}
to the extremal state.  The argument of the exponential operator $D(s)$ consists of a sum over
operators the complement of $H$ in $G$, and the parameters $s_j$ are coordinates desginating points on the
geometrical space of the quotient group $G/H$.  The complex metric can be written~\cite{Chaturvedi87} as
\begin{equation}
  \label{eqn:C_DDDD}
  C_2(j;k) = \sum_{m \neq 0} \langle 0 |[\hat{D}^\dagger(s) \partial_j \hat{D}(s)]^\dagger|m\rangle \langle m |\hat{D}^\dagger(s) \partial_k \hat{D}(s)|0\rangle.
\end{equation}

It is also possible to construct {\it generalized}~\cite{Boiteux73,Roy82} coherent states via acting with the displacement operator on a state which is not extremal, but still an eigenstate of the stability group.  Such states are no longer minimum uncertainty states, but they are interesting, because their time-dependent generalization corresponds~\cite{Roy82} to the averages moving according to classical dynamics, while the full quantum distribution remains rigid.  Other generalizations of the original coherent state formalism include a variety of squeezed states~\cite{Zhang90,Mahran86,Gerry89,Kim89}.

\textcolor{black}{To summarize, our construction of coherent states and the calculation of the quantum geometric tensor in subsections \ref{ssec:Glauber}, \ref{ssec:SU2}, and \ref{ssec:SU11} will proceed via the following steps:
\begin{itemize}
\item  Identify the stability group and its complement.
\item  Identify states which are eigenstates of the operators of the stability group.
\item  Apply the shift operator to eigenstates of the stability group.
\item  Calculate the quantum geometric tensor
\end{itemize}}

\textcolor{black}{For the case of $SU(1,1)$ we will first analyze (subsection \ref{ssec:SU11}) single-valued representations of the group which correspond to the universal covering space of the hyperboloid.  It is also possible to construct projective representations in which the groups are multi-valued.  In subsection \ref{ssec:SU11_pr} we give one example of this.  The reader should note that the group theoretical approach to potential problems is a very colorful subject, whose mathematical intricacies are beyond the scope of this work,  but excellent references exist to quench unsatisfied further curiosity~\cite{Bargmann47,Alhassid83,Alhassid86}.}

\subsection{Uncertainty principle in terms of coherent states}
  
We can derive an uncertainty relation of the Schrödinger type for Lie
group coherent states of the form of Eqs. (\ref{eqn:cs_Glauber}).
These relations will include only the operators of the coset space,
and we show that it remains valid at any point on the geometrical
space $G/H$.  We can write the operator
\begin{equation}
  \label{eqn:DpartialD}
  \hat{D}^\dagger(s) \partial_k \hat{D}(s) = \sum_l A_{kl}(s) \hat{B}_l,
\end{equation}
where $A_{kl}(s)$ is a matrix which depends on the coordinates $s$.
This linear combination includes only operators of the coset space.
Therefore, we can write $C_2(j;k)$ as
\begin{equation}
  C_2(j;k) = \langle 0 |[\hat{D}^\dagger(s) \partial_j \hat{D}(s)]^\dagger \hat{D}^\dagger(s) \partial_k \hat{D}(s)|0\rangle.
\end{equation}
Using Eq. (\ref{eqn:DpartialD}) we can write $C_2(j;k)$ as
\begin{equation}
  C_2(j;k) = \sum_{lm} A_{jl}^\dagger(s) A_{km}(s)\langle 0
  |\hat{B}_l^\dagger \hat{B}_m |0\rangle.
\end{equation}
The general complex variance can be written as
\begin{equation}
  C_2(j;k) = \sum_{lm} A_{jl}^\dagger(s) A_{km}(s)C_2^{(0)}(l;m).
\end{equation}
Taking the determinant leads to
\begin{equation}
  \mbox{Det} [C_2] = \mbox{Det}[A^\dagger(s) A(s)] \mbox{Det}[C_2^{(0)}].
\end{equation}
Since $\mbox{Det}[A^\dagger A]$ is greater than or equal to zero, and we
already know that this is so for $\mbox{Det}C_2^{(0)}$ from the
uncertainty relation, it follows that
\begin{equation}
  \mbox{Det} [C_2] \geq 0,
\end{equation}
alternatively, it follows that the usual uncertainty relation will
hold anywhere in the parameter space.

\subsection{Glauber coherent states}

\label{ssec:Glauber}

For Glauber coherent states the relevant group is the Weyl-Heisenberg group~\cite{Weyl50}, which consists of elements $\{ \hat{I}, \hat{n}, \hat{a}^\dagger, \hat{a}\}$, where $\hat{a}$($\hat{a}^\dagger$) denotes the annihilation(creation) operator, and $\hat{n}=\hat{a}^\dagger \hat{a}$.   The stability group is $\{\hat{I},\hat{n}\}$.  Minimum uncertainty coherent states are generated using the shift operator, as, 
\begin{equation}
  |\alpha \rangle = \exp(\alpha \hat{a}^\dagger - \alpha^* \hat{a}) |0
  \rangle,
\end{equation}
where $\alpha = \alpha_1 + i \alpha_2$ (the complex plane) is the parameter space.  We will also consider generalized Glauber states, constructed from an excited harmonic
oscillator state,
\begin{equation}
  |\alpha \rangle_m = \exp(\alpha \hat{a}^\dagger - \alpha^* \hat{a}) |m
  \rangle.
\end{equation}
Using Eq. (\ref{eqn:C_DDDD}), the result for the quantum geometric tensor is,
\begin{equation}
  C_2 = \begin{pmatrix}
    2 m + 1 & i \\
    -i & 2 m + 1
  \end{pmatrix}.
\end{equation}
For $m>0$ the metric is invertible and there exists a non-trivial
complex metric, however, for $m=0$ the metric is not invertible, and
the geometry is trivial.  \textcolor{black}{Stated differently, coherent states constructed starting from extremal states give rise to a trivial geometry, whereas generalized coherent states are endowed with a non-trivial geometry.}

\subsection{$SU(2)$ (atomic) coherent states}

\label{ssec:SU2}
 
The Lie group of angular momentum $G$ is generated by the angular
momentum Lie algebra, $\mathfrak{g} = \{\hat{J}_x, \hat{J}_y,
\hat{J}_z\}$, which obeys the commutation relations, 
\begin{equation}
  [\hat{J}_i,\hat{J}_j] = i \hat{J}_k,
\end{equation}
where $i,j,k$ denote a cyclic permutation of the coordinates $x,y,z$.
For later use we also define the raising and lowering operators,
\begin{equation}
  \label{eqn:JpJm}
  \hat{J}_\pm = \hat{J}_x \pm i \hat{J}_y.
\end{equation}
A convenient basis is formed by the states $|j,m\rangle$, which
satisfy,
\begin{eqnarray}
  \hat{J}^2 |j,m \rangle &=& j(j+1) |j,m \rangle \\ \nonumber
  \hat{J}_z |j,m \rangle &=&  m |j,m \rangle,
\end{eqnarray}
and for which
\begin{eqnarray}
\label{eqn:jm_SU2}
  j &=& \frac{1}{2},1,\frac{3}{2},2,... \\ \nonumber
  m &=& -j,-j+1,-j+2,...,j-2,j-1,j.
\end{eqnarray}  
The Casimir operator of the group $G$, $\hat{J}^2$, is defined as
\begin{equation}
  \hat{J}^2 = \hat{J}_x^2 + \hat{J}_y^2 + \hat{J}_z^2.
\end{equation}
The stability group $H$ of the group $G$, is the subgroup generated by
rotations about the $z$-axis (subalgebra $\mathfrak{h} = \{ \hat{J}_z
\}$).  The coset space of $G/H$ corresponds to a sphere (which we can parametrize
by angles $\theta,\phi$).

A generalized coherent state ($|\tau \rangle$) can be generated by the $SU(2)$ shift
operator
\begin{equation}
  D(\tau) = \exp(\tau \hat{J}_+)\exp(\beta \hat{J}_z)\exp(-\tau^* \hat{J}_-),
\end{equation}
where 
\begin{eqnarray}
  \tau &=& \tan \frac{\theta}{2} e^{i\phi} \\
  \beta &=& \ln(1 + |\tau|^2), \nonumber
\end{eqnarray}
by acting on a basis state
\begin{equation}
  |\tau \rangle = D(\tau)|j,m\rangle.
\end{equation}
The usual atomic coherent states, which are minimum uncertainty
states, are generated by the above procedure from the extremal
basis functions \textcolor{black}{of which in this case there are two the lower and upper bounds of the series of spherical harmonic solutions} ($m = -j$ or $m=j$).

After some algebra it is possible to show that
\begin{eqnarray}
  \nonumber
  D(\tau)^\dagger \partial_\theta D(\tau) &=& \frac{1}{2}\left[e^{-i\phi} \hat{J}_+ - e^{i\phi} \hat{J}_- \right],  \\
  D(\tau)^\dagger \partial_\phi D(\tau) &=& \frac{-i \sin \theta}{2}\left[e^{-i\phi} \hat{J}_+ + e^{i\phi} \hat{J}_- \right].
  \label{eqn:DdD_atomic}
\end{eqnarray}
Using Eq. (\ref{eqn:C_DDDD})), we obtain the complex metric for the
atomic coherent states,
\begin{equation}
  \label{eqn:C_2_SU2_def}
    C_2 =\frac{1}{2}\begin{pmatrix} j(j+1) - m^2 & - i m \sin \theta \\ i m \sin \theta & (j(j+1) - m^2)\sin^2 \theta \end{pmatrix}.
\end{equation}
Taking the determinant results in,
\begin{equation}
  \mbox{Det}[C_2] = \frac{1}{4}[j(j+1)-m(m+1)][j(j+1)-m(m-1)]\sin^2\theta.
\end{equation}

The PVM together with the BC is recovered by setting $m=-j$ in
Eq. (\ref{eqn:C_2_SU2_def}).  We also find that there are four values
of $m$ for which the determinant vanishes, $m=-j-1,-j,j,j+1$, but two
of these do not contribute, because they are outside the range of $m$
for the representation of the $SU(2)$ group (\textcolor{black}{Eq. (\ref{eqn:jm_SU2})}).  Hence, for arbitrary
values of the coordinates, only the extremal states correspond to a
trivial complex quantum metric.  For the remaining states the metric
is non-trivial, except if $\theta=0,\pi$, which are the north and
south poles of the sphere.

We can consider a mixed quantum classical system of the type
chronicled in Section \ref{sec:BO_approx}.  A $2 \times 2$ inverse
mass tensor has the form,
\begin{equation}
  \label{eqn:Q_tensor}
  Q = \begin{pmatrix} Q'_{11} & Q'_{12} \\ Q'_{12} & Q'_{22} \end{pmatrix} - i \begin{pmatrix} 0 & Q''_{12} \\ -Q''_{12} & 0 \end{pmatrix}
\end{equation} 
leading to a potential term of the form,
\begin{equation}
  \Phi(\theta) = \left(j(j+1) - m^2\right)(Q'_{11} + Q'_{22} \sin^2 \theta) + 2 m Q''_{12} \sin \theta.
\end{equation}
The second term is the contribution due to the imaginary component of
the inverse mass tensor.

\subsection{$SU(1,1)$ coherent states}

\label{ssec:SU11}

$SU(1,1)$ coherent states are generated by the Lie algebra,
$\mathfrak{g} = \{\hat{J}_x, \hat{J}_y, \hat{J}_z\}$, whose members
obey the relations
\begin{eqnarray}
  \label{eqn:SU11_Lie}
  [\hat{J}_x,\hat{J}_y] &=& - i \hat{J}_z, \\ \nonumber
  [\hat{J}_y,\hat{J}_z] &=&   i \hat{J}_x, \\ \nonumber
  [\hat{J}_z,\hat{J}_x] &=&   i \hat{J}_y.
\end{eqnarray}
The raising and lowering operators are defined according to
Eq. (\ref{eqn:JpJm}).  A convenient basis is formed by the states
$|j,m\rangle$, which satisfy,
\begin{eqnarray}
\label{eqn:jmSU11}
  \hat{J}^2 |j,m \rangle &=& j(j+1) |j,m \rangle \\ \nonumber
  \hat{J}_z |j,m \rangle &=&  m |j,m \rangle,
\end{eqnarray}
but here the Casimir operator of the group $G$, $\hat{J}$, is defined
as
\begin{equation}
  \hat{J}^2 = \hat{J}_x^2 + \hat{J}_y^2 - \hat{J}_z^2.
\end{equation}
The algebra of the stability group is $\mathfrak{h} = \{ \hat{J}_z\}$,
the coset space of $G/H$ corresponds to a hyperboloid (parametrized
below by $\rho,\phi$).

The $SU(1,1)$ group has several possible series of quantum numbers, depending on representation\cite{Alhassid86}.  Using the
notation used in Ref. \cite{Alhassid86}, here we list the ones corresponding to the single-valued representations of the group.  There are two discrete
series, $D_j^+$ and $D_j^-$, and two continuous series, $C_k^0$ and
$C_k^{1/2}$.  For $D_j^+$, the quantum number $j$ can take negative integer or
half-integer values, while, $m$ can start from $-j$ and can increase
in integral steps, but unlike the $SU(2)$ series, the $m$ quantum
number is not bounded above.  We have
\begin{eqnarray}
  j &=& -\frac{1}{2},-1,-\frac{3}{2},-2,... \\ \nonumber
  m &=& -j,-j+1,-j+2,...
\end{eqnarray}
The series of $m$ values is only bounded from below.  For the series $D_j^-$, the quantum number $j$ can take negative integer or
half-integer values, while, $m$ can start from $j$ and can decrease in
integral steps,
\begin{eqnarray}
  j &=& -\frac{1}{2},-1,-\frac{3}{2},-2,... \\ \nonumber
  m &=& j,j-1,j-2,...
\end{eqnarray}
The series of $m$ values are only bounded from above.  For the two continuous series,
\begin{eqnarray}
  j &=& -\frac{1}{2} + ik, \\ \nonumber
  m &=& 0,\pm 1, \pm 2, ..., \hspace{.2cm}\mbox{for}\hspace{.2cm} C_k^0, \\ \nonumber
  m &=& \pm \frac{1}{2},\pm \frac{3}{2} , ..., \hspace{.2cm}\mbox{for}\hspace{.2cm} C_k^{1/2},
\end{eqnarray}
where $k$ is a real number, such that $k>0$.  

A generalized $SU(1,1)$ coherent state ($|\tau \rangle$) is generated
by the $SU(1,1)$ shift operator,
\begin{equation}
  D(\tau) = \exp(\tau \hat{J}_+)\exp(\beta \hat{J}_z)\exp(-\tau^* \hat{J}_-),
\end{equation}
where 
\begin{eqnarray}
  \tau &=& -\tanh \frac{\rho}{2} e^{-i\phi} \\
  \beta &=& \ln(1 - |\tau|^2), \nonumber
\end{eqnarray}
by acting on a basis state $|j,m\rangle$
\begin{equation}
  |\tau \rangle = D(\tau)|j,m\rangle.
\end{equation}

After some algebra it is possible to show that
\begin{eqnarray}
  \nonumber
  D(\tau)^\dagger \partial_\rho D(\tau) &=& -\frac{1}{2}\left[e^{-i\phi} \hat{J}_+ -
    e^{i\phi} \hat{J}_- \right], \\ 
  D(\tau)^\dagger \partial_\phi D(\tau) &=& \frac{i}{2}\sinh(\rho)  \left[e^{-i\phi} \hat{J}_+ + e^{i\phi} \hat{J}_- \right].
  \label{eqn:DdD_su11}
\end{eqnarray}
Using Eq. (\ref{eqn:C_DDDD})), we obtain the complex metric for the
hyperbolic coherent states,
\begin{equation}
  \label{eqn:C_2_def}
    C_2 =\frac{1}{2}\begin{pmatrix} \left[-j(j+1) + m^2\right] & - i m\sinh \rho \\ i m\sinh \rho & \left[-j(j+1) + m^2\right]\sinh^2 \rho \end{pmatrix}.
\end{equation}
Taking the determinant results in,
\begin{equation}
  \label{eqn:DetC2_SU11}
  \mbox{Det}[C_2] = \frac{1}{4}[j(j+1)-m(m+1)][j(j+1)-m(m-1)]\sinh^2\rho.
\end{equation}

This determinant is zero if $\rho=0$.  It is also possible to get zero
for particular values of the quantum numbers.  The determinant
of the complex quantum metric is zero for the extremal states in each
of the series $D_j^+$ ($m=-j$) and $D_j^-$ ($m=j$).    For the two continuous series
there is no way for the determinant of the complex quantum metric to
be zero.  We note in passing that there exists~\cite{Alhassid86} also
a supplementary series for the $SU(1,1)$ group.  This series does not
contribute to the solution of the relevant Laplace equation (defined
on a hyperboloid), since the series 
$D_j^+$, $D_j^-$, $C_k^0$, and $C_k^{1/2}$ 
together form a complete set in which any function can be
expanded.  For this supplementary series,
\begin{eqnarray}
  -\frac{1}{2}< j < 0, \\ \nonumber
  m = 0,\pm 1, \pm 2,...
\end{eqnarray}
The determinant of the complex quantum metric is never zero.

We also consider a mixed quantum classical system with inverse mass
tensor of the form given in Eq. (\ref{eqn:Q_tensor}), the potential
term will be of the form,
\begin{equation}
 \Phi(\rho) = (-j(j+1) + m^2)(Q'_{11} + Q'_{22} \sinh^2 \rho) + 2 m Q''_{12} \sinh \rho.
\end{equation}
Again, the second term is the contribution due to the imaginary
component of the inverse mass tensor.

\subsection{Projective representations of $SU(1,1)$ and two coupled oscillators as an example}

\label{ssec:SU11_pr}

\textcolor{black}{Group spaces can have single-valued and multi-valued representations.  The former correspond to the universal covering space of the group, whereas the latter to projective representations.  An intuitive example is the case of the one-dimensional translation group, which is single-valued, and its group space is the one-dimensional line.  The group of uniaxial rotations correspond to a {\it projective} representation of the one-dimensional translation group, but in this case the representation is no longer single valued, and the group space will be the circle.  The one-dimensional translation group is known as the universal covering group of the group of uniaxial rotations.  Analogously, the one-dimensional line is the universal covering space of the circle.   Exactly this situation occurs for $SU(1,1)$.  In addition to the single-valued representations studied in the previous subsection, $SU(1,1)$ also has {\it projective} representations, which correspond to multi-valued group representations.   
}

The Lie algebras are the same (Eq. (\ref{eqn:SU11_Lie})), because locally the universal covering space and the group space of the multivalued represesntations is the same.  The form of the quantum metric is also the same as of the single-valued representations, given by Eq. (\ref{eqn:C_2_def}), what changes are the series of quantum numbers.  We state without proof that the {\it projective discrete} representation corresponds to the the series,
\begin{equation}
\label{eqn:proj_dscrt}
  m = -j,-j+1,...,
\end{equation}
for $j<0$, $j$ real.  In this case, $\mbox{Det}[C_2]$ is zero for
$m=-j$.  The projective continuous representation is characterized by
two numbers,
\begin{eqnarray}
  j = -\frac{1}{2} + i \delta \hspace{.5cm}\delta=\mbox{real}>0,\\ \nonumber
  m_0 = \mbox{real},\hspace{.5cm} 0 \leq m_0 < 1.
\end{eqnarray}
The spectrum of $J_z$ in this case is
\begin{equation}
  m = m_0, m_0 \pm 1,m_0 \pm 2,...
\end{equation}

As an example we can consider the case of a pair of coupled
oscillators, a system which has been studied in
Ref. \onlinecite{Mahran86,Gerry89,Kim89,Levay93}.  We define
\begin{equation}
\label{eqn:2osc}
  \hat{J}_+ = \frac{1}{2} \hat{a}^\dagger \hat{a}^\dagger, \hat{J}_- = \frac{1}{2} \hat{a} \hat{a}, \hat{J}_z = \frac{1}{2}\left(\hat{a}^\dagger \hat{a} + \frac{1}{2} \right),
\end{equation}
where $\hat{a}^\dagger$($\hat{a}$) are bosonic creation(annihilation)
operators.  It can easily be shown that the operators $\hat{J}_+$,
$\hat{J}_-$, and $\hat{J}_z$ satisfy the $SU(1,1)$ commutation rules, \textcolor{black}{showing that the Lie algebra is the same as before}.
\textcolor{black}{The Casimir operator is also unchanged, meaning that Eq. (\ref{eqn:jmSU11}) is still valid.}
After some algebra we find, though, that the Casimir operator, using the definitions in Eq. (\ref{eqn:2osc}) becomes,
\begin{equation}
\label{eqn:3p16}
  \hat{J}^2 = -\frac{3}{16}\hat{I},
\end{equation}
where $\hat{I}$ denotes the identity.   Solving Eq. (\ref{eqn:jmSU11}) results in two possible $j$ values,
\begin{equation}
 j = -\frac{1}{4},-\frac{3}{4}.
\end{equation}
\textcolor{black}{The essential point is that Eq. (\ref{eqn:3p16})  does not fit into any of the sequences given in subsection \ref{ssec:SU11} ($D_j^+$, $D_j^-$, $C_k^0$, and $C_k^{1/2}$ 
) even though the Lie algebra of the generators is unchanged.   However, it can be accommodated by the sequence given in Eq. (\ref{eqn:proj_dscrt}). }
\begin{equation}
  m = \frac{1}{4},\frac{5}{4},... \hspace{.5cm} m = \frac{3}{4},\frac{7}{4},...
\end{equation}
The two series are each distinct projective discrete representations,
and their direct sum is the representation of the two-oscillator
system.  The series corresponding to $j = -\frac{1}{4}$($j =
-\frac{3}{4}$) are the odd(even) photon squeezed vacuum
states~\cite{Gerry89} (see also
Refs. \onlinecite{Mahran86,Gerry89,Kim89,Levay93}).
In such a system, characterized by a direct
sum, there are two instances where the geometry becomes trivial,
$m=\frac{1}{4}$ and $m=\frac{3}{4}$.

\section{Conclusion}

\label{sec:cnclsn}

In this paper we investigated the complex geometric tensor of quantum
state manifolds.  We showed that the imaginary part of this tensor,
the Berry curvature, gives rise to a potential term in mixed
quantum-classical systems obeying the Born-Oppenheimer approximation,
if the inverse mass tensor has an imaginary component.  

From the scalar product of two quantum states a generating function
can be written, which generates, in principle, all orders of quantum
fluctuations.  The second cumulant (variance) corresponds to the
complex quantum metric, higher order cumulants correspond to higher
order geometric quantities, the skew is related to the affine
connection, the kurtosis gives the complex analog of the four-index
Riemann curvature tensor.  Requiring the determinant of the complex
quantum metric to be positive definite gives generalized uncertainty
relations.  Our calculations for Lie group coherent states led to
trivial geometries for the usual coherent states, when they are
generated from an extremal state, however for other cases, the
determinant of the complex metric can be nontrivial.  Of particular
interest is the representation of the $SU(1,1)$ group formed by
coupling a pair of harmonic oscillator creation and annihilation
operators, whose representation consists of the direct sum of two
projective representations of the group.  In this case the extremal
states of both direct sums give zero for the determinant of the
complex quantum metric, meaning the geometry is trivial.  Again, for
other states the geometry is non-trivial.

We envision a variety of interesting further studies
  based on the formalism we presented.  In quantum optics, the
  properties of squeezed coherent states are measured via various
  correlation functions of the number
  operator~\cite{Mahran86,Gerry89,Kim89}, which are four operator
  products of creation and annihilation operators.  Another
  interesting direction would be to study time-dependent coherent
  states to connect the underlying geometry with the dynamics.

\section*{Acknowledgments}

BH was supported by the National Research, Development and Innovation
Fund of Hungary within the Quantum Technology National Excellence
Program (Project Nr. 2017-1.2.1-NKP-2017-00001) and by the
BME-Nanotechnology FIKP grant (BME FIKP-NAT).  PL was supported by the
Ministry of Culture and Innovation and the National Research,
Development and Innovation Office within the Quantum Information
National Laboratory of Hungary (Grant No. 2022-2.1.1-NL-2022-00004).
\\

\appendix

\section{Moments and cumulants}

In this appendix, we derive the moments and cumulants from the scalar
product of two quantum states and compare them to relations known for
ordinary moments and cumulants used in statistics.  First we review
the definition of ordinary moments and cumulants.

Given a multi-variate normalized probability distribution:
\begin{eqnarray}
  P(x_1,...,x_N) \geq 0, \forall x_i; \hspace{3cm}\\
  \int_{-\infty}^\infty...\int_{-\infty}^\infty d x_1 ... d x_N P(x_1,...,x_N) = 1, \nonumber
\end{eqnarray}
The generating (or characteristic) function is defined as,
\begin{equation}
  f(k_1,...,k_N) = \int d x_1 ... d x_N e^{i \sum_{j=1}^N k_j x_j} P(x_1,...,x_N).
\end{equation}
Moments are obtained by taking derivatives of $f(k_1,...,k_N)$.  The
first three moments can be written,
Cumulants are logarithmic derivatives of the characteristic function.
\begin{eqnarray}
  \mathcal{M}_1(l) &=&  \frac{1}{i}\frac{\partial}{\partial k_l} f(k_1,...,k_N)|_{\bf k = 0}, \\ \nonumber
   &=&  \langle x_l \rangle, \\ \nonumber
  \mathcal{M}_2(l,m) &=&  \frac{1}{i^2}\frac{\partial^2}{\partial k_l\partial k_m} f(k_1,...,k_N)|_{\bf k = 0}, \\ \nonumber
   &=&  \langle x_l x_m\rangle, \\ \nonumber
  \mathcal{M}_3(l,m,n) &=&  \frac{1}{i^3}\frac{\partial^3}{\partial k_l\partial k_m\partial k_n} f(k_1,...,k_N)|_{\bf k = 0}, \\ \nonumber
  &=&  \langle x_l x_m x_n\rangle.
\end{eqnarray}
The first three cumulants are defined as
\begin{eqnarray}
  \mathcal{C}_1(l) &=&  \frac{1}{i}\frac{\partial}{\partial k_l} \ln f(k_1,...,k_N)|_{\bf k = 0}, \\ \nonumber
  \mathcal{C}_2(l,m) &=&  \frac{1}{i^2}\frac{\partial^2}{\partial k_l\partial k_m} \ln f(k_1,...,k_N)|_{\bf k = 0}, \\ \nonumber
  \mathcal{C}_3(l,m,n) &=& \frac{1}{i^3}\frac{\partial^3}{\partial k_l\partial k_m\partial k_n} \ln f(k_1,...,k_N)|_{\bf k = 0},
\end{eqnarray}
or in terms of moments, they can be written,
\begin{eqnarray}
  \label{eqn:CC_MM}
  \mathcal{C}_1(l) &=&  \mathcal{M}_1(l), \\ \nonumber
  \mathcal{C}_2(l,m) &=& \mathcal{M}_2(l,m) - \mathcal{M}_1(l)\mathcal{M}_1(m), \\ \nonumber
  \mathcal{C}_3(l,m,n) &=& \mathcal{M}_3(l,m,n) - \mathcal{M}_2(l,m)\mathcal{M}_1(n) \\ & & - \mathcal{M}_2(l,n)\mathcal{M}_1(m) - \mathcal{M}_2(m,n)\mathcal{M}_1(l) \nonumber \\ & & + 2 \mathcal{M}_1(l) \mathcal{M}_1(m) \mathcal{M}_1(n). \nonumber
\end{eqnarray}
In the formalism used in our work the role of the characteristic
function is played by the scalar product $S(s',s) = \langle \Psi(s')|
\Psi(s) \rangle$ (Eq. (\ref{eqn:Sss})).  There are two sets of
variables, the primed ones, $s' = s'_1,...,s'_n$, associated with the
bra vector of the scalar product, and the unprimed ones, $s =
s_1,...,s_n$, associated with the ket.  Derivatives of either one can
be used to define the analog of a moment.  There are two types of
first moments, for which we introduce the following notation,
\begin{eqnarray}
  \mathcal{M}_1(\_;l) &=& (i \partial_l) S(s',s)|_{s' = s} \\ \nonumber
  \mathcal{M}_1(l;\_) &=& (-i \partial_l') S(s',s)|_{s' = s}.
\end{eqnarray}
The underscore $\_$ is used when no derivatives are taken for a
particular set of variables.  For the first moments, it holds that
\begin{eqnarray}
 \mathcal{M}_1(\_;l) = \mathcal{M}_1(l;\_) = \mathcal{M}_1(l) \\ \nonumber = - \beta_l,
\end{eqnarray}
see Eq. (\ref{eqn:C_2_def_c}) for the definition of $\beta_l$.
Depending on whether derivatives are applied to the primed or unprimed
variables, there are three different types of second moments.  We
write them as,
\begin{eqnarray}
  \mathcal{M}_2(\_;kl) &=& (i \partial_k)(i \partial_l) S(s',s)|_{s' = s} \\ \nonumber
  \mathcal{M}_2(k;l) &=& (-i \partial_k')(i \partial_l) S(s',s)|_{s' = s} \\ \nonumber
  \mathcal{M}_2(kl;\_) &=& (-i \partial_k')(-i \partial_l') S(s',s)|_{s' = s}.
\end{eqnarray}
$\mathcal{M}_2(k;l)$ corresponds to $\gamma_{kl}$ in
Eq. (\ref{eqn:C_2_def_c}).  As for third moments, there are four
distinct types, but we will only write the two that are useful in
establishing the third order Christoffel symbol (section
\ref{sec:3index}),
\begin{eqnarray}
  \mathcal{M}_3(j;kl) &=& (-i \partial_j')(i \partial_k)(i \partial_l) S(s',s)|_{s' = s} \\ \nonumber
  \mathcal{M}_3(jk;l) &=& (-i \partial_j')(-i \partial_k')(i \partial_l) S(s',s)|_{s' = s} 
\end{eqnarray}

Cumulants are obtained in a similar way, by applying derivatives to
$\ln S(s',s)$.  The first order cumulants can be written,
\begin{eqnarray}
  \mathcal{C}_1(\_;l) &=& (i \partial_l) \ln S(s',s)|_{s' = s} \\ \nonumber
  \mathcal{C}_1(l;\_) &=& (-i \partial_l') \ln S(s',s)|_{s' = s}.
\end{eqnarray}
the second ones as,
\begin{eqnarray}
  \mathcal{C}_2(\_;kl) &=& (i \partial_k)(i \partial_l) \ln S(s',s)|_{s' = s} \\ \nonumber
  \mathcal{C}_2(k;l) &=& (-i \partial_k')(i \partial_l) \ln S(s',s)|_{s' = s} \\ \nonumber
  \mathcal{C}_2(kl;\_) &=& (-i \partial_k')(-i \partial_l') \ln S(s',s)|_{s' = s}.
\end{eqnarray}
and the third order ones we use here as,
\begin{eqnarray}
  \mathcal{C}_3(j;kl) &=& (-i \partial_j')(i \partial_k)(i \partial_l) \ln S(s',s)|_{s' = s} \\ \nonumber
  \mathcal{C}_3(jk;l) &=& (-i \partial_j')(-i \partial_k')(i \partial_l) \ln S(s',s)|_{s' = s}.
\end{eqnarray}

In the text it was shown that cumulants which have at least one
derivative as a function of each set of parameters (the primed and
unprimed) are gauge invariant.  Moments are not gauge invariant.  It
can also be shown that cumulants can be expressed in terms moments,
and the expressions look similar to Eqs. (\ref{eqn:CC_MM}), with
appropriate modifications.  For first order cumulants, it holds that,
\begin{equation}
  \mathcal{C}_1(\_;l) = \mathcal{C}_1(l;\_) = \mathcal{M}_1(l).
\end{equation}
For the second order cumulants, we have,
\begin{eqnarray}
  \mathcal{C}_2(\_;kl) &=&  \mathcal{M}_2(\_;kl) - \mathcal{M}_1(\_;k)\mathcal{M}_1(\_;l)\\ \nonumber
  \mathcal{C}_2(k;l) &=&  \mathcal{M}_2(k;l) - \mathcal{M}_1(k;\_)\mathcal{M}_1(\_;l)\\ \nonumber
  \mathcal{C}_2(kl;\_) &=&  \mathcal{M}_2(kl;\_) - \mathcal{M}_1(k;\_)\mathcal{M}_1(l;\_).
\end{eqnarray}
The indices on both sides of the equation ``keep their sides of the
semi-colon'', otherwise the second cumulants obey the same relation as
for ordinary statistical cumulants (Eq. (\ref{eqn:CC_MM})).  For the
third order cumulants, we have,
\begin{eqnarray}
  \mathcal{C}_3(j;kl) &=&  \mathcal{M}_3(j;kl) - \mathcal{M}_2(j;k)\mathcal{M}_1(\_;l)\\ \nonumber
     & &   - \mathcal{M}_2(j;l)\mathcal{M}_1(\_;k) - \mathcal{M}_1(j;\_)\mathcal{M}_2(\_;kl)\\ \nonumber
     & &   + 2\mathcal{M}_1(j;\_)\mathcal{M}_1(\_;k)\mathcal{M}_1(\_;l), \\ \nonumber
  \mathcal{C}_3(jk;l) &=&  \mathcal{M}_3(jk;l) - \mathcal{M}_1(j;\_) \mathcal{M}_2(k;l)\\ \nonumber
     & &   - \mathcal{M}_1(k;\_)\mathcal{M}_2(j;l) - \mathcal{M}_2(jk;\_)\mathcal{M}_1(\_;l)\\ \nonumber
     & &   + 2\mathcal{M}_1(j;\_)\mathcal{M}_1(k;\_)\mathcal{M}_1(\_;l).
\end{eqnarray}
Again, the indices maintain ``their sides of the semicolon''.  Further
simplifications can be made, by using $\mathcal{M}_1(l)$, but we
wanted to emphasize the pattern of the placement of indices.

\end{document}